\documentclass[runningheads,fleqn]{svmult}

\usepackage{makeidx}   
\usepackage{graphicx}  
\usepackage{subeqnar}  
\usepackage{multicol}  
\usepackage{physproc}  
\makeindex             
\newcommand{\Ham}{{\cal H}}
\newcommand{\eqref}[1]{(\ref{eq:#1})}
\newcommand{\figref}[1]{Fig.~\ref{fig:#1}}
\newcommand{\secref}[1]{Sect.~\ref{sec:#1}}
\begin{document}
\title*{Quantum World-line Monte Carlo Method with Non-binary Loops and Its Application}
\titlerunning{Non-binary Loop Algorithm}
\author{Kenji Harada}
\institute{Graduate School of Informatics, Kyoto University, Kyoto 606-8501, Japan}
\maketitle

\begin{abstract}
  A quantum world-line Monte Carlo method for high-symmetrical quantum
  models is proposed.  Firstly, based on a representation of a
  partition function using the Matsubara formula, the principle of
  quantum world-line Monte Carlo methods is briefly outlined and a new
  algorithm using non-binary loops is given for quantum models with
  high symmetry as SU($N$).  The algorithm is called non-binary loop
  algorithm because of non-binary loop updatings.  Secondary, one
  example of our numerical studies using the non-binary loop updating
  is shown.  It is the problem of the ground state of two-dimensional
  SU($N$) anti-ferromagnets. Our numerical study confirms that the
  ground state in the small $N (\le 4)$ case is a magnetic ordered Neel
  state, but the one in the large $N (\ge 5)$ case has no magnetic
  order, and it becomes a dimer state.
\end{abstract}

\section{Introduction}
\label{sec:intro}
The quantum world-line Monte Carlo (QMC) method has a long
history. It has been used in many numerical studies of
condensed-matter physics since it first proposed in
1970's \cite{Suzuki1976}. However QMC simulations suffered as the
result of some defects in conventional QMC algorithms. For example,
near a critical point, or at low temperatures, the correlation time in
samples became extremely long, thus we could not calculate the
canonical ensemble average with accuracy. There were same problems in
the Monte Carlo simulation of a classical system.  Fortunately, a new
algorithm was proposed by Swendsen and Wang, which solves such
problems in classical systems. It is called cluster algorithm. The
idea of the cluster algorithm can apply to quantum cases. Indeed,
Evertz, Lana and Marcu first proposed a new QMC algorithm based on
the cluster algorithm, which called loop
algorithm \cite{EvertzLM1993}. The loop algorithm evolves into recent
powerful QMC algorithms (See reviews \cite{Evertz2003} and
\cite{KawashimaH2004}).

In the present article, we will focus to the loop algorithm and the
generalization for high-symmetrical quantum models.  In
\secref{world-line-configuration}, a useful representations of a
partition function for recent QMC algorithms will be derived. In
\secref{qmc}, the principle of QMC method and the detail of the loop
algorithm will be reviewed. In \secref{non-binary-loop-algorithm}, our
generalization of the loop algorithm to a non-binary loop updating
will be proposed. In \secref{application}, we will show one example of
our numerical studies with the non-binary loop updating: the problem
of the ground state of the two-dimensional SU($N$) quantum
anti-ferromagnets.

\section{World-line Representations Based on the Matsubara Formula}
\label{sec:world-line-configuration}

A density operator has all informations of a quantum system in an
environment. In particular, if the system is in the heat-bath whose
inverse temperature is $\beta$, the density operator $\rho(\beta)$ is
the exponential operator of the system Hamiltonian as $\rho(\beta) =
\exp(-\beta \Ham)$, where $\Ham$ is the Hamiltonian of a quantum
system.

The Hamiltonian usually consists of two parts $\Ham_0$ and $V$. Here
$\Ham_0$ and $V$ denote diagonal and non-diagonal operators,
respectively. The $V$ generally represents a perturbation part of the
Hamiltonian. Then the non-trivial part of the density operator,
$\rho^\prime(\beta)$, satisfies a special Bloch equation in which
there is explicitly only the $V$ operator:
\begin{equation}
  \label{eq:Bloch}
  \frac{\D \rho^\prime(\beta)}{\D \beta} = -V(\beta)\rho^\prime(\beta) \qquad
  \left(\rho(\beta)\equiv \E^{-\beta(\Ham_0 +V)} \equiv \E^{-\beta\Ham_0}\rho^\prime(\beta)\right),
\end{equation}
where $V(t)$ denotes the interaction picture of $V$ operator as $V(t)
\equiv \E^{t\Ham_0}V\E^{-t\Ham_0}$. The solution of \eqref{Bloch}
formally satisfies an integral equation as
\begin{equation}
  \label{eq:integral}
  \rho^\prime(\beta) = I - \int_0^\beta \D t V(t)\rho^\prime(t),
\end{equation}
where an integral variable $t$ is called {\it imaginary time}, because
the Bloch equation is related to the Schr\"odinger equation through
analytic continuation.

The solution of \eqref{integral} is written as a series of multiple
integrals of the product of interaction pictures of $V$:
\begin{equation}
  \label{eq:matsubara-1}
  \rho^\prime(\beta) = 
    I - \int_0^\beta \D t_1 V(t_1) + \int_0^\beta \D t_2 \int_0^{t_2} \D t_1 V(t_2)V(t_1) - \cdots
    .
\end{equation}
This representation of a density operator is known as the Matsubara
formula.  The order of interaction pictures $V(t)$ is in descending
order.  Since the perturbation part $V$ is usually the sum of local
interaction Hamiltonians $V_b$, the density operator is a series of
multiple integrals of the product of interaction pictures of $V_b$:
\begin{equation}
  \label{eq:matsubara-2}
  \rho(\beta) = \E^{-\beta\Ham_0}\sum_{n=0}^\infty\mathop{\sum}_{b_n,\cdots,b_1}\mathop{\int}_{\beta \ge t_n \ge \cdots \ge t_1 \ge 0}
  \D t_n \cdots \D t_1 (-1)^n V_{b_n}(t_n)\cdots V_{b_1}(t_1),
\end{equation}
where $V=\sum_b V_b$.

The matrix element of operator $AB$ is equal to the summation of
products of each matrix elements of $A$ and $B$.
\begin{equation}
  \langle \Psi_A \vert AB \vert \Psi_B \rangle = \sum_{\{\vert \psi \rangle\} \mbox{:basis}}
  \langle \Psi_A \vert A \vert \psi \rangle \langle \psi \vert B \vert \Psi_B \rangle,
\end{equation}
where the set $\{\vert \psi \rangle\}$ is an orthonormal basis in the quantum
state space. Therefore, inserting an orthonormal basis $\{ \vert \psi
\rangle \}$ between two operators of integrand in the Matsubara
formula, the partition function $Z$ can be represented as a multiple
integral with respect to three kinds of variables, $\psi_i,\ b_i$ and
$t_i$:
\begin{eqnarray}
  Z &\equiv& \mathrm{Tr}\rho(\beta)\\
  &=& \sum_{n=0}^\infty
  \mathop{\sum}_{\psi_n,\cdots,\psi_1(\psi_{n+1}=\psi_1)}\mathop{\sum}_{b_n,\cdots,b_1}\mathop{\int}_{\beta \ge t_n \ge \cdots \ge t_1 \ge 0} W_n(\{\psi_i\},\{b_i\},\{t_i\}),  \label{eq:partition-1}\\
  &&W_n(\{\psi_i\},\{b_i\},\{t_i\}) \equiv \E^{-\beta\Ham_0(\psi_1)}\prod_{i=1}^{n}
  \langle \psi_{i+1} \vert \{-V_{b_i}(t_i)\D t_i\} \vert \psi_i \rangle\label{eq:weight-world-line},
\end{eqnarray}
where $\Ham_0(\psi_1) \equiv \langle \psi_1 \vert \Ham_0 \vert \psi_1 \rangle$.

\begin{figure}[hbt]
\begin{center}
\includegraphics[width=0.5\textwidth]{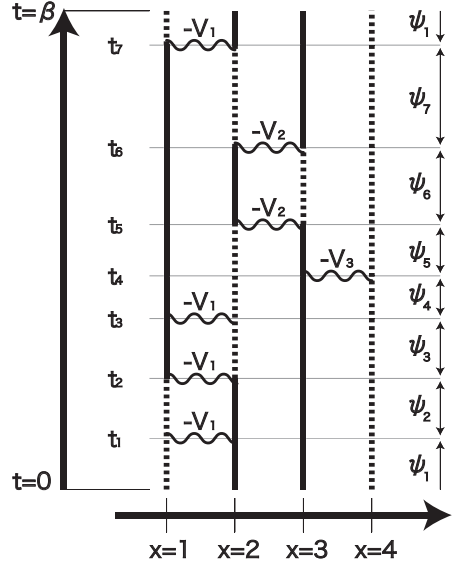}
\end{center}
\caption[]{A graphical representation of a world-line configuration
  for an $s=1/2$ Heisenberg anti-ferromagnet model on a four-sites
  chain. Solid and dotted lines denote spin states $+\frac12$ and
  $-\frac12$, respectively. A position of a waiving line corresponds
  to the status of a variable set $(b_i, t_i)$.}
\label{fig:world-line}
\end{figure}

The status of a variable set $(\{\psi_i\}, \{b_i\}, \{t_i\})$ is
called world-line configuration, because it can be represented as a
set of world-lines. Figure \ref{fig:world-line} shows a world-line
configuration of an $s=1/2$ Heisenberg anti-ferromagnet (HAF) model on
a four-sites chain.
The X axis refers the position of spin sites, from 1 to 4, and the Y
axis refers the imaginary time, from 0 to $\beta$. In this example,
the Hamiltonian consists of local interaction Hamiltonians $V_1$,
$V_2$ and $V_3$. The $V_b$ is the Hamiltonian of the HAF interaction
between sites $b$ and $b+1$. Then a waving line is drawn at the
position corresponding to the status of $(b_i, t_i)$. The vertical
line denotes the status of $\psi_i$ between two $V_b$. In the $s=1/2$
case, we need two line types, for example, solid and dotted lines
which denote up and down spin states of an $s=1/2$ spin,
respectively. In this way, this figure is in one-to-one correspondence
with the world-line configuration. Usually, the waving line is called
vertex and the position at where a local spin configuration changes is
called kink. Since we will only show examples of quantum spin models
in the present article, the status of variables $\psi_i$ is called
spin configuration.

From \eqref{partition-1}, the function $W_n$ can be regarded as the
Boltzmann weight of a world-line configuration. Therefore, using the
Matsubara formula, the partition function can be transformed from
quantum to classical one.

\section{Quantum World-line Monte Carlo Method}
\label{sec:qmc}
Using the same technique for the partition function, we can get a
similar multiple integral representation of a canonical ensemble
average of an observable A.
\begin{equation}
	\langle A \rangle \equiv \frac{\mathrm{Tr} A \rho(\beta)}{Z} = \sum_{(\{\psi_i\}, \{b_i\}, \{t_i\})}
		A(\psi_1) \frac{W_n(\{\psi_i\},\{b_i\},\{t_i\})}{Z},
		\label{eq:canonical-1}
\end{equation}
where $A(\psi_1) \equiv \langle \psi_1 \vert A \vert \psi_1
\rangle$. Here the symbol $\sum_{(\{\psi_i\}, \{b_i\}, \{t_i\})}$
denotes the multiple integral for variables $(\{\psi_i\}, \{b_i\},
\{t_i\})$.  In other words, we sum up with respect to all world-line
configurations.

Since $\sum_{(\{\psi_i\}, \{b_i\}, \{t_i\})} W_n = Z$, the function
$W_n/Z$ in \eqref{canonical-1} can be regarded as the probability of a
world-line configuration.  Therefore, if we can sample a world-line
configuration with the probability $W_n/Z$, the canonical ensemble
average $\langle A \rangle$ can be calculated as the average of
samples of the function $A(\psi_1)$. This is the principle of the QMC
method.

Many algorithms have been proposed to make a sample of world-line
configurations.  However the conventional algorithm before the loop
algorithm suffered from some problems.  In general, since the
probability distribution of world-line configurations is complex, a
sampling algorithm of world-line configurations is based on a Markov
process.  Samples in a Markov process are usually correlated for some
time steps which is called correlation time. Unfortunately the
correlation time in the conventional sampling algorithm becomes very
large near a critical point and at low temperatures, because the size
of an updating unit of a world-line configuration is fixed when the
correlation length of a quantum system increases. Therefore it is
difficult to calculate a canonical ensemble average of a quantum
system at interesting points using the conventional algorithm.

\subsection{Loop Algorithm}
The loop algorithm is one of recent developed QMC algorithms.  It uses
a geometrical object which called loop and updates a world-line
configuration with loops globally. For some important quantum models,
we can prove that the size of loop in the loop algorithm is equal to
the correlation length of the quantum model. This is the reason why
the loop algorithm will avoid problems in QMC simulations. In fact,
the loop algorithm has some good properties as follows. The first good
property is that the correlation time in samples is very short at a
critical point and low temperatures. The second one is that various
types of world-line configurations can be sampled naturally, which
called grand canonical sampling. In practice, it is difficult to
change the value of an order parameter without artificial techniques
in the conventional algorithm.

\begin{figure}[hbt]
\begin{center}
\includegraphics[width=1\textwidth]{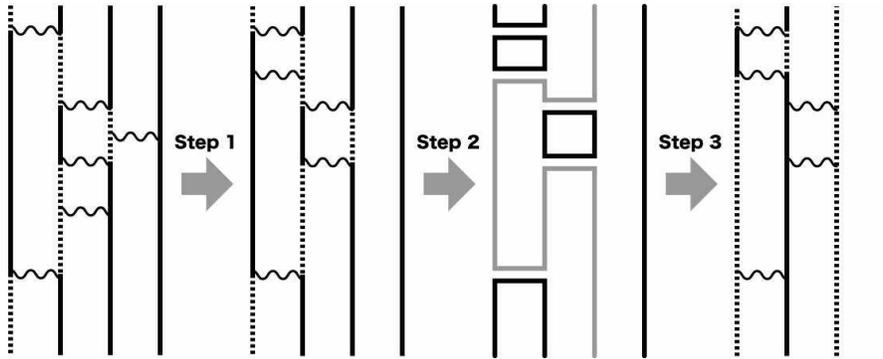}
\end{center}
\caption[]{A loop updating for an $s=1/2$ Heisenberg anti-ferromagnet
  model on a four-sites chain.  Solid and dotted line denote spin
  states $+\frac12$ and $-\frac12$, respectively. The gray loop in
  the third sub-figure is only flipped.}
\label{fig:loop-steps-1}
\end{figure}

The procedure of a loop algorithm consists of three steps. Figure
\ref{fig:loop-steps-1} illustrates these steps for an $s=1/2$ HAF model
on a four-sites chain.
The first step is to update vertexes.  In this step, the number of
vertexes is changed and those positions are moved, but the spin
configuration is fixed. We separately update vertexes and the spin
configuration.  The second step is to decompose spin variables into
loops. In \figref{loop-steps-1} they are decomposed into five
loops. The third step is to flip each loop with a probability $1/2$.
The flipping of a loop is to change the sign of spin variables on the
loop. In the quantum $s=1/2$ spin model, a basis of a quantum state
space can be made of the direct product of two states, up and down
states, on each spin site. Thus a spin variable takes only two values
as $\pm 1/2$ and it can be flipped. Although the gray loop in
\figref{loop-steps-1} is only flipped, the spin configuration is
drastically changed to the right one. Since there are great
differences between the initial and the final world-line
configurations, the correlation time of a loop algorithm is very
short.

Since the sampling distribution defined by the weight of a world-line
configuration is very complex and the partition function is unknown in
advance, we use the Markov process whose equilibrium distribution of
configurations is equal to the desired sampling distribution. In order
to make the Markov process, it is enough to satisfy the detail balance
condition for any two configurations A and B in the Markov process:
\begin{equation}
W({\mathrm A}) P({\mathrm A} \to {\mathrm B}) = W({\mathrm B}) P({\mathrm B} \to {\mathrm A}),
\label{eq:detail}
\end{equation}
where $W(\cdot)$ is the weight function of a desired distribution and
$P({\mathrm A} \to {\mathrm B})$ is the transfer probability from A to
B. In other words, the density of random walkers is controlled through
their transfer probabilities which balance with the desired sampling
distribution. In the loop algorithm, two updatings for vertexes and
spin configurations are alternately done and each transfer probability
satisfies the detail balance condition.

Firstly, the updating of vertexes is considered. If a vertex $V_b$ is
inserted at an imaginary time $t$ between $t_{k+1}$ and $t_k$
($t_{k+1} \ge t \ge t_k$), the weight of a new world-line
configuration $W_{n+1}$ is equal to the present one ($W_n$ in
\eqref{weight-world-line}) multiplied by a matrix element of $V_b$
with $\D t$: $\langle \psi_{k+1}\vert \{ -V_b(t)\D t\}
\vert\psi_{k+1}\rangle$. The ratio of two weights of pre- and
post-inserted a vertex $V_b$ between the imaginary time $t$ and $t +
\D t$ is proportional to the infinitesimal $\D t$. Therefore, from
\eqref{detail}, the accept probability of the inserting of the vertex
is proportional to the infinitesimal $\D t$. Such stochastic process
is called Poisson process.  Scanning a world-line configuration with
the Poisson process, we can insert vertexes under the detail balance
condition. The thing to do next is to remove a vertex.  If a vertex is
not on a kink, the ratio of weights between pre- and post-removed
configurations is infinite. From \eqref{detail}, we can always remove
a vertex on a non-kink.  On the other hand, if a vertex is on a kink,
the ratio of them is exactly zero. Thus it can not be can removed. In
summary, the stochastic inserting process of vertexes is a Poisson
process whose intensity is equal to a matrix element of a vertex. And
the removing a vertex must be done if and only if the vertex is not on
a kink.

This updating procedure of vertexes by a Poisson process is applicable
to various quantum models, and it is employed in various QMC
algorithms. However, on the contrary, it is difficult to find an
efficient updating of a spin configuration because of the interaction
of spin variables thorough vertexes. In particular, the weight of
random generated spin configurations is usually equal to zero, because
the matrix element of each vertex often takes a value 0.  Thus we need
an ingenious updating procedure.

Fortunately, it has been found for some important quantum models. In
such cases, the interaction Hamiltonian takes the special form which
is called delta operator. The delta operator is defined as the matrix
element taking only a value 0 or 1.  For example, the HAF interaction
Hamiltonian is equal to a special delta operator $\hat{\Delta}(g_H)$
divided by 2 with a special unitary transformation:
\begin{equation}
-\Ham_{ij}^\mathrm{HAF} \equiv -U^{-1} \left( \vec{S}_i \cdot \vec{S}_j - \frac14 \right) U= \frac12 \hat{\Delta}(g_H).
\end{equation}
Furthermore, the condition of the matrix element taking a value 1 is
as follows:
\begin{equation}
\langle \sigma_i^\prime\sigma_j^\prime \vert \hat{\Delta}(g_H) \vert \sigma_i \sigma_j \rangle
\equiv \left\{\begin{array}{cl}
1 & (\mbox{if} \ \sigma_i + \sigma_j = \sigma_i^\prime + \sigma_j^\prime =  0)\\
0 &(\mbox{otherwise})
\end{array}\right.,
\end{equation}
where $\sigma_i (\sigma_j)$ denotes a spin state on a site $i(j)$ and
it takes an eigenvalue of an $s=1/2$ z-direction spin operator as
$\sigma_i \equiv \pm \frac12$.

The condition of taking a value 1 for a delta operator can be often
represented by a graph. If a delta operator of an interaction
Hamiltonian has the graphical representation, we can construct the
loop algorithm on the quantum model. For example, the condition of the
$\hat{\Delta}(g_H)$ operator is simply represented by a horizontal
graph in \figref{loop-steps-2} (\textbf{a}).  The horizontal dotted
line in \figref{loop-steps-2} (\textbf{a}) denotes the necessary
condition: the sum of two connected spin variables must be equal to
zero. From the definition, a matrix element of a delta operator takes
only a value 0 or 1. Therefore the value of a vertex weight is not
changed, as long as a modification of a spin configuration matches the
graphical representation. In the horizontal graph, the modification of
the two connected spin variables is only restricted, thus disconnected
spin variables are not related to each other: $\sigma_i$ and
$\sigma_i^\prime$. Furthermore, the flipping of two connected spin
variables together is always allowed, because it satisfies the
condition and the weight is not changed. Therefore, we can always flip
each set of connected spin variables independently.

In summary, the procedure of updating a spin configuration consists of
three steps. In the first step each vertex is replaced with a
corresponding graph. The second step is to decompose spin variables
into a set which defined by graphs.  Since a set of connected spin
variables always becomes a loop for the $s=1/2$ HAF model, this
updating algorithm is called loop algorithm. The third step is to flip
each loop or set randomly with a probability $\frac12$.  The loop takes
one of two states, because the spin variable takes only a value
$\pm\frac12$ for the $s=1/2$ HAF model. Thus we call it {\it binary
  loop}.

The loop updating was very successful, because the size of loop is
related to the physical correlation length of a quantum model. And by
the loop algorithm, many important results were numerically found for
various quantum models.

Although we showed only a quantum $s=1/2$ spin model, the loop
updating is also applicable to quantum large spin models. For such
cases, Kawashima and Gubernatis proposed a mapping to a quantum
$s=1/2$ spin model which called split-spin
technique \cite{KawashimaG1994}. The idea is that an $s=m$ spin
operator is replaced by the sum of $2m$ $s=1/2$ spin operators with a
permutation operator. Since they are isomorphic, we can simulate a
quantum $s=m$ model as a multi quantum $s=1/2$ model with binary loop
updatings.  However, since the size of the new quantum space becomes
large, we need more memories and complex treatments. In the next
section, we will propose a new loop updating which acts on an original
quantum space directly.

\section{Non-binary Loop Algorithm}
\label{sec:non-binary-loop-algorithm}
The symmetry of a model is related to the graphical representation of
a vertex.  For example, the HAF model has the SU(2) symmetry and the
HAF interaction term is represented as the binary horizontal
graph. Fortunately, there are such special graphical representations
in high-symmetrical cases.

\begin{figure}[hbt]
\begin{center}
\includegraphics[width=1\textwidth]{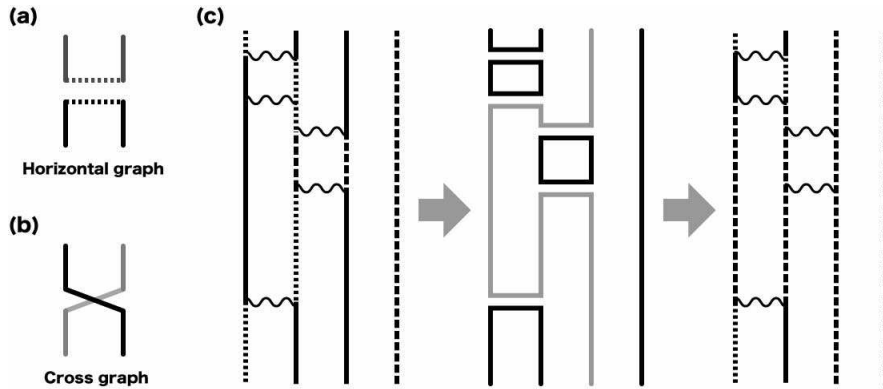}
\end{center}
\caption[]{Graphs for various types of vertexes and a non-binary loop
  updating for an $s=1$ BLBQ model on a four-sites chain. ({\bf a})
  Horizontal graph. ({\bf b}) Cross graph. ({\bf c}) The non-binary
  loop updating for an $s=1$ BLBQ model at $\theta=-\frac{\pi}{2}$.
  Solid, dashed, and dotted lines denote the spin states, +1, 0, and
  -1, respectively. The gray loop in the second sub-figure is only
  changed.}
\label{fig:loop-steps-2}
\end{figure}

The $s=1$ bi-linear bi-quadratic (BLBQ) model is interesting
theoretically and experimentally \cite{BIQ}, because it has some
integrable points for a one-dimensional case and it is related to the
effective model of Na atoms in an optical
lattice \cite{GreinerMEHB2002}. The Hamiltonian is as follows:
\begin{equation}
\Ham_{ij}^{\mathrm{BLBQ}} \equiv (\cos\theta) \vec{S}_i\cdot\vec{S}_j + (\sin\theta)(\vec{S}_i\cdot\vec{S}_j)^2.
\end{equation}
The BLBQ model has SU(3) symmetry at special points, for example,
$\theta=-\frac{\pi}{2}$ and $-\frac{3\pi}{4}$. The BLBQ interaction
terms at SU(3) points have special graphical representations whose
states are non-binary as 0 and $\pm 1$. For example, the BLBQ
interaction term at the SU(3) point $\theta=-\frac{\pi}{2}$ becomes a
delta operator and the condition taking a value 1 is same to that in
an $s=1/2$ HAF case, except for a spin variable takes three values,
$\pm 1$ and $0$. As a result, we can use the same horizontal graph in
\figref{loop-steps-2} (\textbf{a}) for the BLBQ interaction term.
However, the loop, which defined by the SU(3) horizontal graph, takes
three states, not binary states.

We can generalize the horizontal graph in the SU($N$) case. In the
SU($N$) case, since the spin variable takes $N$ values ($\sigma_i =
\frac{(-N+1)}{2} , \cdots, \frac{(N-1)}{2}$), the number of loop
states which defined by the SU($N$) horizontal graph is $N$. Thus we
call it {\it non-binary loop}. Another type of graph can also be
generalized in the SU($N$). For example, we found the generalized
cross graph $g_C$ in \figref{loop-steps-2} (\textbf{b}) which
corresponds to the other SU(3) point $\theta=-\frac{3\pi}{4}$:
\begin{equation}
\langle \sigma_i^\prime\sigma_j^\prime \vert \hat{\Delta}(g_C) \vert \sigma_i \sigma_j \rangle
\equiv \left\{\begin{array}{cl}
1 & (\mbox{if} \ \sigma_i - \sigma_j^\prime = \sigma_i^\prime - \sigma_j =  0)\\
0 &(\mbox{otherwise})
\end{array}\right..
\end{equation}
It is easy to check that generalized graphs have the SU($N$)
symmetry.

The loop defined by the SU($N$) graph has $N$ possible states of equal
weight, because the exchange of any two states corresponds to one of
SU($N$) transformations. Therefore we can choose one of $N$ possible
states for each loop with equal probability.

In summary, the procedure of the loop updating using the SU($N$) graph
is similar to that of the conventional loop algorithm with the binary
graph.  In the first step each vertex is replaced by a corresponding
SU($N$) graph.  And the second step is to decompose spin variables
into a set which defined by the SU($N$) graphs. In the third step we
choose one of $N$ possible states for each set with equal probability.
The only difference between the conventional and new loop algorithms
is the number of possible loop states.

The loop algorithm with non-binary loops is called {\it non-binary
  loop algorithm} \cite{KawashimaH2004}.  It is very useful to study
high-symmetrical quantum models, because of the simplicity.  Without a
split-spin technique, we can directly simulate the SU($N$) model in
the original quantum space.  The relation between the conventional and
new loop algorithms is similar to that between Swendsen--Wang
algorithms for Ising and Potts models.

\section{Application: Ground States of Two-Dimensional SU($N$) Quantum
  Anti-ferromagnets}
\label{sec:application}
Some quantum models already have been simulated with the non-binary
loop algorithm.  In the present section, we will focus on the
numerical study of ground states of two-dimensional SU($N$) quantum
anti-ferromagnets.

The existence of a short-range resonating valence bond (RVB)
spin-liquid is a very interesting problem for low-dimensional quantum
systems.  A RVB spin-liquid exhibits a finite gap for spin-excitations
and it has only short-range order and it does not break any lattice
symmetry.  Although the RVB spin-liquid may be related to the
mechanism of the copper-oxide superconductors, the ground state of the
two-dimensional $s=1/2$ HAF model, which is an effective quantum spin
model of copper-oxide superconductor, is not a RVB state, but it has a
magnetic order which called Neel order.  As the spin $s$ is decreased,
the quantum fluctuation decreases the Neel order, but the ground state
remains ordered even for $s=1/2$. Another route to increase quantum
fluctuations is in the symmetry of quantum models. We studied the
SU($N$) quantum anti-ferromagnets which is one of generalizations of
the HAF model and which has the higher symmetry than SU(2).

Our Hamiltonian of SU($N$) quantum anti-ferromagnets is defined by the
generator of SU($N$) algebra as the HAF model:
\begin{eqnarray}
\Ham_{ij}^{\mathrm{SU(}N\mathrm{)}} &\equiv& \frac{1}{N}\sum_{\alpha,\beta=1}^{N} J_\beta^\alpha(i)J^\beta_\alpha(j),\\
\left[J_\beta^\alpha(i), J^\mu_\nu(j) \right] & = & 
\delta_{ij} \left( \delta_\nu^\alpha J_\beta^\mu(i)-\delta_\beta^\mu J_\nu^\alpha(i) \right),
\end{eqnarray}
where $J_\beta^\alpha(i)$ is an SU($N$) generator on site $i$.  In the
SU(2) case, this Hamiltonian is equal to the HAF model.  In our
anti-ferromagnetic SU($N$) model, the representation of SU($N$)
generator is a $N$-dimensional matrix which called fundamental
representation and the matrix on one sub-lattice is conjugate to that
on the other sub-lattice. In this representation, the model can be
expressed in terms of SU(2) spins with $s=(N-1)/2$. If the SU($N$)
Hamiltonian applies to the $\vert n m \rangle$ state, the result
equals the superposition of all $\vert l\bar{l}\rangle$ states divided
by $N$:
\begin{equation}
  \Ham_{ij}^{\mathrm{SU(}N\mathrm{)}} \vert nm\rangle = -\delta_{n\bar{m}} \left(\frac{1}{N}\sum_{l=-s,\cdots,s}\vert l\bar{l}\rangle \right)
  \equiv -\delta_{n\bar{m}} \vert (ij) \rangle. \quad (\bar{l} \equiv -l),
\end{equation}
where $\vert nm \rangle$ denotes the simultaneous eigenstate of
z-direction SU(2) spin operators on sites $i$ and $j$. For
convenience, the special superposition state $\vert (ij) \rangle$ will
be called $(ij)$-bond state in the following.

The $(12)$-bond state is a ground state of a two-sites case. And it
has no magnetic order, because it is the superposition of different
magnetic ordered states.  In a four-sites case, a bond state $\vert
(12) (34) \rangle$ is an eigenstate of
$\Ham_{12}^{\mathrm{SU(}N\mathrm{)}}$ and
$\Ham_{34}^{\mathrm{SU(}N\mathrm{)}}$ whose eigenvalue is -1.
However, if the interaction Hamiltonian does not match the covering of
dimers over sites in a bond state, the state converges to a
zero-eigenstate when $N$ becomes infinity:
\begin{equation}
\Ham_{23}^{\mathrm{SU(}N\mathrm{)}}\vert (12) (34) \rangle = -\frac1N\vert (41) (23) \rangle \to 0 \quad
(N \to \infty).
\end{equation}
Therefore, in the large $N$ limit, the ground state maximizes the number
of nearest-neighbor bonds, because the eigenvalue of a whole
Hamiltonian is minus of the number of nearest-neighbor bonds.  In
short, in the large $N$ limit, due to strong quantum fluctuations, the
ground state has no magnetic order.  However, in a one-dimensional
chain, the number of maximum bond state is only two \cite{Affleck1985}.
Since they break the translational symmetry, they are not a RVB state,
which called dimer state. Read and Sachdev studied the ground state of
SU($N$) models on a two-dimensional square lattice
theoretically \cite{ReadS1990}. Their conclusion is also that the
ground state on a two-dimensional square lattice is a dimer state in
the large $N$ limits.

In the $N=2$ case, the ground state is a magnetic ordered Neel state.
On the other hand, in the large $N$ limit, it has no magnetic order
and it becomes a dimer state by Read and Sachdev's prediction. However
the intermediate $N$ cases remain problems.  The numerical study by
Santoro {\it et al.} suggested a RVB spin-liquid ground state in the
$N=4$ case \cite{Santoro1999}.  The model of Santoro {\it et al.} is
equal to our SU(4) model and it also appears as a special point in
coupled spin-orbital models. Santoro {\it et al.} used the Green
function Monte Carlo method in their numerical simulations. Due to the
computational times by their numerical method, the size of their
simulated lattices was limited to only $L=12$.

Since the SU($N$) interaction Hamiltonian is equal to the delta
operator $\hat{\Delta}(g_H)$ represented by the non-binary horizontal
graph, this model can be simulated for various $N$ and large square
lattices by a non-binary loop algorithm.  Indeed, we explored the
system size $L$ up to 128 and $N$ up to 8 \cite{HaradaK2003}.  Our
simulations have been performed at low enough temperatures to be
effectively in the ground state.  The lowest temperature for $N=4$ and
$L=128$, for example, is $\frac{1}{128}$. 

In order to search the order of a ground state, we have measured two
quantities which are related to magnetic and dimer orders,
respectively. The one is a static structure factor $S(\pi,\pi)$ of
staggered magnetization $m_s$ and the other is a dimer static
structure factor $S^\mathrm{D}(\pi,0)$:
\begin{equation}
\frac{S(\pi,\pi)}{L^2} \equiv \langle (m_s)^2 \rangle
\qquad \qquad \quad
\left(m_s \equiv \frac{1}{L^d} \sum_r (-1)^r S^z_r\right)
\end{equation}
and
\begin{equation}
\frac{S^\mathrm{D}(\pi,0)}{L^2} \equiv \langle (d_{(\pi,0)})^2 \rangle
\ \quad \qquad
\left(d_{(\pi,0)} \equiv \frac{1}{L^d} \sum_r (-1)^{r_x} S^z_rS^z_{r+x}\right).
\end{equation}
If an order exists, the related structure factor divided by $L^2$
converges to a finite value in the large-$L$ limits, but if the order is absent, it decreases to zero
as $1/L^2$. 

\begin{figure}[hbt]
\begin{center}
\includegraphics[width=0.9\textwidth]{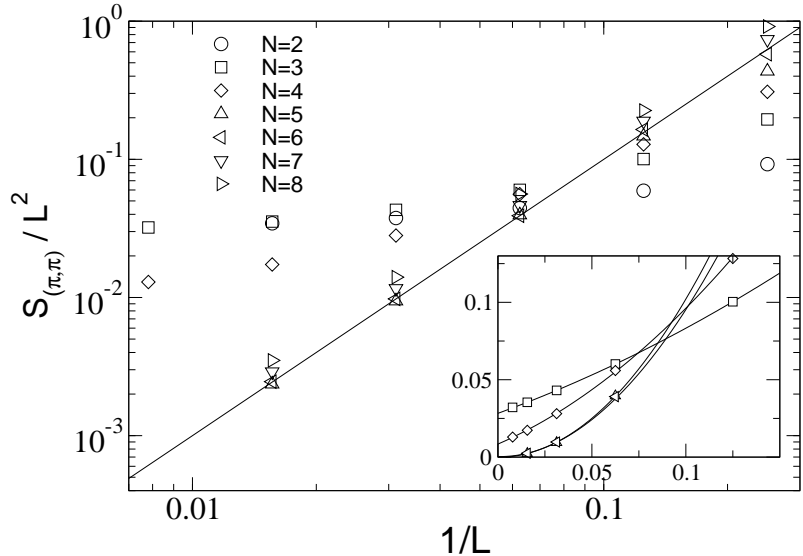}
\end{center}
\caption[]{Static structure factors $S_{(\pi,\pi)}$ for $2 \le N \le
  8$.  The straight line representing the power law $L^{-2}$ is drawn
  for comparison.  Estimated statistical errors are not shown, because
  they are equal to or smaller than the symbol size.  The inset
  presents the data for $N=3, 4, 5, 6$ in the linear scale, together with
  the best fitting curves obtained by the method of least squares.
  (Adopted from Harada and Kawashima \cite{HaradaK2003})  }
\label{fig:ss}
\end{figure}

Figure \ref{fig:ss} shows a clear evidence for a Neel order in $N \le
4$, because the spin structure factor divided by $L^2$ converges to a
finite value for $N \le 4$.  Therefore the ground state is not a RVB
spin-liquid even in the $N=4$ case. On the contrary, in the $N \ge 5$
cases, the spin structure factor divided by $L^2$ decreases as
$1/L^2$.  Therefore there is no Neel order in $N \ge 5$.
The disagreement of our conclusion for the $N=4$ case with that of
Santoro {\it et al.} is not in the raw numerical data. Their data are
only in the $L \le 12 $ region. Therefore, the disagreement is solely
due to the small system sizes studied in \cite{Santoro1999}.

\begin{figure}[hbt]
\begin{center}
\includegraphics[width=0.9\textwidth]{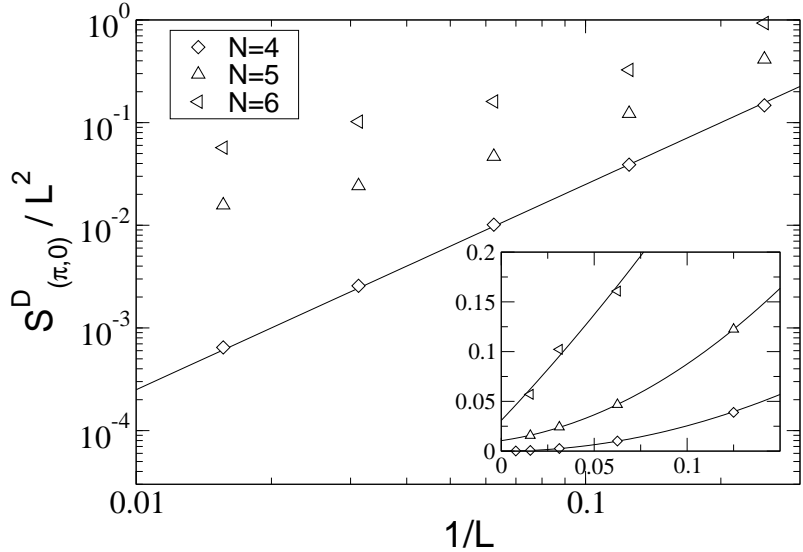}
\end{center}
\caption[]{The ${\bf k} = (\pi, 0)$ dimer structure factors
  $S_{(\pi,0)}^\mathrm{D}$ for $N=4, 5, 6$ in logarithmic
  scale. The inset is the linear-scale plot.  The solid lines in the
  inset are the best fitting curves of least squares based on the
  $L\ge 8$ data.  (Adopted from Harada and Kawashima
  \cite{HaradaK2003})  }
\label{fig:sd}
\end{figure}

On the other hand, \figref{sd} shows a clear evidence for a dimer
order in $N \ge 5$. While the dimer structure factor divided by $L^2$
decreases to zero in $N = 4$, it converges to a finite value in $N \ge
5$.  Therefore the dimer order exists in $N \ge 5$ and the ground
state is not a RVB spin-liquid state. It is consistent with the
theoretical prediction by Read and Sachdev in the large $N$ limits.

In summary, from QMC simulations for large square lattices by the loop
algorithm with non-binary loops, we confirmed a direct transition from
the Neel ground state ($N\le 4$) to the dimer ground state ($N \ge 5$)
for two-dimensional SU($N$) quantum anti-ferromagnets. These
comprehensive numerical studies can be effectively done by the loop
algorithm with non-binary loops.

\section{Conclusion}
We have introduced the QMC method with non-binary updatings for
quantum models with high symmetry. For example, using graphical
representations of the interaction Hamiltonian, we can easily simulate
a high-symmetrical quantum model by a loop algorithm with non-binary
loops. We have only focused to the loop algorithm, but the idea of a
non-binary updating for quantum models with high symmetry can be
combined with the other algorithms as the worm or directed loop. We
have shown one example for using a non-binary loop algorithm. It is
the problem of the ground state of two-dimensional SU($N$) quantum
anti-ferromagnets. Our conclusion is that our SU($N$) models have no
RVB spin-liquid ground state.

\section*{Acknowledgments}
The author is grateful to N.~Kawashima and Y.~Okabe for stimulating
conversations and useful comments.

\end{document}